\documentclass[journal,onecolumn]{IEEEtran}
\usepackage{authblk}
\usepackage{graphicx}
\usepackage{kotex}
\usepackage{amsfonts}
\usepackage{amssymb}
\usepackage{setspace}
\usepackage{cite}
\ifCLASSINFOpdf

\else

\fi

\begin{document}

\title{Wireless Map-Reduce Distributed Computing with Full-Duplex Radios and Imperfect CSI}

\author{Sukjong~Ha$^{1}$, Jingjing~Zhang$^{2}$, Osvaldo~Simeone$^{2}$,      and~Joonhyuk~Kang$^{1}$
}
\affil{$^{1}$ KAIST, School of Electrical Engineering, South Korea\\
$^{2}$ King's College London, Centre for Telecommunications Research, London, United Kingdom\\
$^{1}$ sj.ha@kaist.ac.kr, jhkang@ee.kaist.ac.kr, $^{2}$ \{jingjing.1.zhang, osvaldo.simeone\}@kcl.ac.uk}

\maketitle
\begin{abstract}
Consider a distributed computing system in which the worker nodes are connected over a shared wireless channel. 
Nodes can store a fraction of the data set over which computation needs to be carried out, and a Map-Shuffle-Reduce protocol is followed in order to enable collaborative processing. 
If there is exists some level of redundancy among the computations performed at the nodes, the inter-node communication load during the Shuffle phase can be reduced by using either coded multicasting or cooperative transmission. 
It was previously shown that the latter approach is able to reduce the high-Signal-to-Noise Ratio communication load by half in the presence of full-duplex nodes and perfect transmit-side Channel State Information (CSI). 
In this paper, a novel scheme based on superposition coding is proposed that is demonstrated to outperform both coded multicasting and cooperative transmission under the assumption of imperfect CSI.
\end{abstract}
\vspace{-2mm}
\section{Introduction}
\label{sec:intro}
\vspace{-2mm}
Distributed computing platforms are the current method of choice for the implementation of many computational tasks, such as learning algorithms \cite{R1}. 
A standard distributed computing framework is Map-Shuffle-Reduce.
Under this protocol, nodes first “map” the assigned data to some Intermediate Values (IVs) through the computation of given functions; then, IVs are ''shuffled'' among the nodes; and finally nodes produce their final result by “reducing” the relevant IVs. 
An important performance bottleneck for these systems is the communication load caused by the shuffling of IVs among the participating computing nodes \cite{R2,R3,R4,R5}. 

It was recently observed that the communication load can be reduced if the computations carried out at the nodes in the Map phase have some degree of redundancy, in the sense that IVs computed at a node are also computed at other nodes. 
In the original works \cite{R2,R3,R4,R5}, nodes are connected by noiseless multicast channels. 
The availability of redundant IVs is leveraged to create coded multicasting opportunities, whereby the signal multicast by one node provides useful information for a number of other nodes that is proportional to the computing redundancy. 
A significant number of follow-up works has offered various refinements of this idea \cite{R6,R7,R8,R9}.

While the approaches reviewed above leverage the availability of redundant IVs at the receiver end of a multicast link, an alternative solution arises when the computing nodes are connected over a shared wireless channel. 
This scenario may be of interest, for instance, for distributed computing platforms in Internet-of-Things applications. 
On a wireless channel, the presence of common IVs can be leveraged to create cooperative transmission opportunities. 
Based on this idea, reference \cite{R10} proposed a cooperative Zero-Forcing (ZF) precoding strategy.
This approach was shown to outperform coded multicasting in the presence of full-duplex nodes and under the assumption that perfect Channel State Information (CSI) is available to design the precoding matrices.

In this paper, a novel scheme based on superposition coding is proposed that is demonstrated to outperform both coded multicasting and cooperative transmission under the assumption that imperfect, or outdated, CSI is available for precoding design. 
As in \cite{R10}, analysis is carried out by focusing on a high-Signal-to-Noise Ratio (SNR) measure of the inter-node communication load in the Shuffle phase. 
The proposed approach reduces to coded multicasting \cite{R5} when CSI is completely unreliable and to cooperative ZF precoding \cite{R10} when CSI is perfect. 
This work contributes to a recent line of work that has demonstrated the advantages of superposition coding in the absence of perfect CSI \cite{R11,R12}.

The rest of the paper is organized as follows. 
Section 2 describes the system model and the performance criterion.
In Section 3, we review and analyze the two reference schemes studied in \cite{R5} and \cite{R10}. 
Section 4 presents the proposed superposition coding-based scheme, and Section 5 presents some results and discussion.

\textbf{Notation}: For any integer $P$ and $J$, we define the set $[P] \doteq \{1,2,\cdots,P\}$, and the set $\{A_{j}\}_{j=1}^{J} \doteq \{A_{1}, \cdots, A_{J}\}$
We define $|\mathcal{A}|$ as the cardinality of set $\mathcal{A}$.
We also define the symbol $\doteq$ to denote an exponential equality: we write $f(P) \doteq P^{\alpha}$ if $\lim_{P \rightarrow \infty} \log(f(P))/\log(P) = \alpha$ holds.
\section{System Model and operation}
\vspace{-1mm}
\subsection{System Model}
As illustrated in Fig. 1, we consider a distributed computing system, in which $K$ full-duplex nodes communicate over a shared wireless channel with the aim of carrying out a computing task.
In particular, the nodes need to evaluate $Q$ functions $\mathcal{F} = \{f_{1},f_{2}, \cdots, f_{Q}\}$ on the input data defined by files $w_{1}, \cdots, w_{n} \in \mathbb{F}_{2^{L}}$ of $L$ bits.
Each function $f_{q}$ is represented by following the standard Map-Reduce formulation as\vspace{-1mm} 
\begin{equation}
f_{q}(w_{1}, \cdots, w_{N}) = h_{q}(g_{q,1}(w_{1}), \cdots, g_{q,N}(w_{N})),\vspace{-1mm} 
\end{equation}
where each Map function $g_{q,n}$ maps the input file $w_{n}$ to an IV $a_{q,n} = g_{q,n}(w_{n}) \in \mathbb{F}_{2^{F}}$ of $F$ bits; and the reduce function $h_{q}$ maps the $N$ IVs $\{a_{q,1}, \cdots, a_{q,N}\}$ to the output value $f_{q}(w_{1}, \cdots, w_{N}) = h_{q}(a_{q,1}, \cdots, a_{q,N}) \in \mathbb{F}_{2^{B}}$ of $B$ bits. 
Each node has a storage capacity of $\mu N$ files, with $\mu \in [0,1]$ being the fractional storage capacity.
In order to enable distributed computation, a network controller (see Fig. 1) shares part of the data to each node while meeting the nodes' storage constraint.
The computation is carried out in a distributed manner by leveraging wireless communication on the shared channel.
As it will be discussed, the process of computing the output functions over the $N$ files is divided into Map, Shuffle, and Reduce phases.
At the end of the process, all outputs of the functions in  $\mathcal{F}$ must be available at some of the nodes for collection by the network controller. 
Specifically, each node $k$ is assigned to compute a subset $\mathcal{F}_{k} \subseteq \mathcal{F}$ of $Q/K$ functions, such that $\bigcup_{k \in [K]} \mathcal{F}_{k} = \mathcal{F}$.

The channel connecting the nodes is assumed to be flat fading, so that the received signal at node $i$ is given as\vspace{-3mm}
\begin{equation}
y_{i} = \sum_{k=1, k \neq i}^{K}h_{k,i}x_{k} + n_{i},\vspace{-3mm}
\end{equation}
where $h_{k,i} \sim \mathcal{CN}(0,1)$ is the  complex channel coefficient between node $k$ and node $i$, for $k$, $i \in [K]$; $x_{k}$ is the transmitted signal from node $k$, which is subject to the power constraint $\mathbb{E}[|x_{k}|^{2}] \leq P$; and $n_{i}(t) \sim \mathcal{CN}(0,1)$ is the additive Gaussian noise at node $i$.
As reflected by (2), as in \cite{R10}, we assume that each node is capable of full-duplex communication, i.e., each node can transmit and receive simultaneously.

Unlike the model in \cite{R10}, we do not assume the availability of perfect CSI at the network controller.
Rather, nodes accurately estimate their channels and forward them to the network controller.
The network controller designs the transmission schedule and beamforming vectors for all nodes by using the received CSI, and transmits its decisions to the nodes.
Due to delays caused by transmission and processing, the CSI available at the network controller is assumed to be outdated with respect to the channel coefficients in (2) realized during communication.
We model the difference between the outdated CSI $\{\hat{h}_{k,i}\}$ and the actual channel realization $\{h_{k,i}\}$ in (2) by writing the mean square error between $h_{k,i}$ and $\hat{h}_{k,i}$ as \vspace{-2mm}
\begin{equation}
\mathbb{E}[|h_{k,i} - \hat{h}_{k,i}|^{2}] \doteq P^{-\alpha},
\end{equation}
for some $\alpha \geq 0$.
This model has been widely adopted in order to study the impact of imperfect CSI in the high-SNR regime (see, e.g., \cite{R13}).
In this regime, the case $\alpha = 0$ is equivalent to having no CSIT, while the case $\alpha = 1$ yields a negligible CSI error.
Note that CSI at the receiver side during transmission is assumed to be accurate, given that each communication can include pilot symbols.\vspace{-3mm}

\begin{figure}[!t]
\centering
\includegraphics[width=5in]{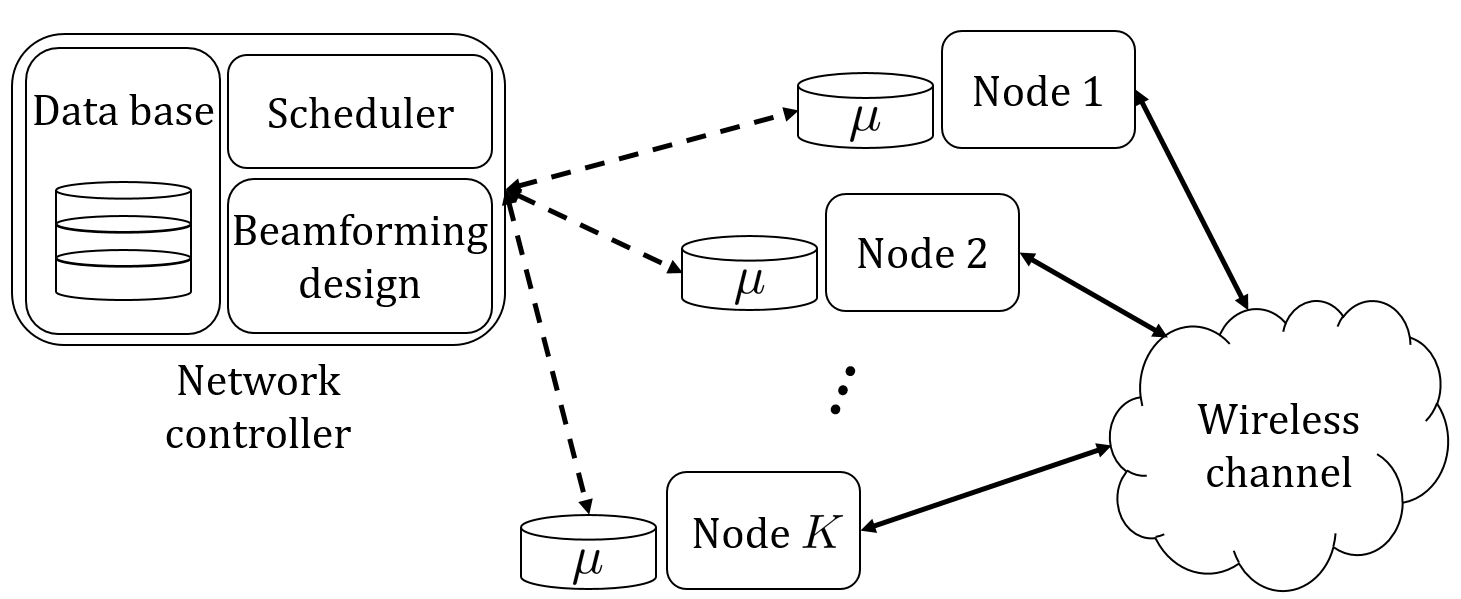}
\caption{Wireless Map-Reduce distributed computing system with $K$ full-duplex radio nodes, each able to store a fraction $\mu$ of the input data, and a network controller.}
\label{fig_system}
\vspace{-3mm} 
\end{figure}

\subsection{Computing and Communication}
\vspace{-1mm}
The three phases of operation of the system are as follows.

1) \emph{Map phase}: In the Map phase, files are first assigned to the nodes by the network controller, and we denote the set of files that are stored at the node $k$ as $\mathcal{M}_{k}$, $\mathcal{M}_{k} \subseteq \{w_{i}\}_{i=1}^{N}$, for $k \in [K]$.
Due to the storage capacity of each node, we have the constraint $|\mathcal{M}_{k}| \leq \mu N$ for all $k$.
In the Map phase, each node computes the IVs for the files that are stored at the node.
In particular, node $k$ computes the $|\mathcal{M}_{k}|Q$ IVs in the set $\mathcal{A}_{k} = \{g_{q,n}(w_{n}) : q \in [Q], n \in \mathcal{M}_{k}\}$.
Accordingly, the computation load of the system is the total number of computed IVs by the nodes over the total number of distinct IVs, i.e., $\sum_{k=1}^{K}|\mathcal{M}_{k}|Q/NQ = \mu K$.

2) \emph{Shuffle phase}: 
As discussed, each node $k$ is assigned to compute the $Q/K$ functions in the set $\mathcal{F}_{k}$.
In order to compute any assigned output function $f_{q} \in \mathcal{F}_{k}$, each node $k$ needs the set of IVs $\left\{a_{q,n} : f_{q} \in \mathcal{F}_{k}, n \in [N]\right\}$.
To this end, the set of IVs that node $k$ needs to receive from other nodes in the Shuffle phase is given as \vspace{-1mm} 
\begin{equation}
\mathcal{B}_{k} = \left\{a_{q,n} : f_{q} \in \mathcal{F}_{k}, n \notin \mathcal{M}_{k}\right\}.\vspace{-1mm} 
\end{equation}
In the Shuffle phase, the $K$ nodes exchange IVs through the wireless channel link by using a transmission of duration $T$ as measured in channel uses.

3) \emph{Reduce phase}: After the Shuffle phase, each node $k$ computes the assigned output functions in $\mathcal{F}_{k}$.\vspace{-2mm}

\subsection{Performance Criterion}
\vspace{-1mm}
As the performance criterion of interest, we adopt the Normalized Communication Load (NCL) required in Shuffle phase in order to successfully carry out the computation.
The NCL is a high-SNR measure of the total average communication time $\mathbb{E}[T]$ needed for the Shuffle phase.
In a manner similar to the Normalized Delivery Time (NDT) \cite{R14}, the average time $\mathbb{E}[T]$ is normalized by time needed in high SNR to transmit all $NQ$ IVs on an interference-free point-to-point link.
Since the high-SNR capacity of such a link is $\log P$ bits per channel use, and given that the total number of IV bits is $NQF$, the NCL for a given scheme is defined as the limit\vspace{-2mm} 
\begin{equation}
\delta(\mu) = \lim_{P \rightarrow \infty}\frac{\mathbb{E}[T]}{NQF/\log(P)}.\vspace{-2mm}
\end{equation} 
We note that this measure is equivalent to the communication load used in \cite{R10}.

\section{Baseline schemes}
In this section, we describe two reference schemes.
The first is a direct application of the coded multicasting in \cite{R5} to the wireless channel at hand. 
The second scheme is the one-shot linear precoding strategy studied in \cite{R10}. 
For the latter scheme, we extend the analysis of \cite{R10} in order to account for outdated CSI (3) at the network controller.
\vspace{-1mm}
\subsection{Coded Multicasting}
\label{subsec:CM}

In \cite{R5}, a coded distributed computing scheme is proposed for a system in which every node is connected to all other nodes over an ideal multicasting links.
In the Map phase, each file is stored at $\mu K$ nodes, hence fully using the available storage capacity of the nodes.
This is done as follows.
For $N$ large enough, we can write the number of files as $N={K \choose \mu K}\eta$, for some $\eta \in \mathbb{N}^{+}$.
The files are equally divided into ${K \choose \mu K}$ batches of size $\eta$, with each batch indexed by a subset $S \subseteq [K]$ of size $\mu K$.
Each node $k$ stores the files in the batch indexed by $S$ if $k \in S$. 

In the Shuffle phase, for each node $k$, the set $\mathcal{B}_{k}$ of required IVs has cardinality $|\mathcal{B}_{k}| = NQ(1-\mu)/K$, since each node can compute a fraction $\mu$ of total number $NQ/K$ of required IVs.
Each one of the required IVs is available at $\mu K$ nodes.
The scheme hence splits each required IV into $\mu K$ sub-IVs, each to be sent by one of the mentioned $\mu K$ nodes.
Specifically, each node multicasts a coded XORed message obtained from $\mu K$ sub-IVs, each destined to a different node.
This is done so that each receiving node has already computed all coded sub-IVs other than the desired sub-IV.
This allows $\mu K$ sub-IVs to be delivered in a single transmission.
Assuming that transmitters operate using time sharing, the NCL can be computed as indicated in the next lemma.

\emph{Lemma 1}: 
For storage capacity $\mu \in \{1/K, 2/K, \cdots, 1\}$ and $\alpha \in [0,1]$, the NCL of coded multicasting is given as
\vspace{-2mm}
\begin{equation}
\delta_{CM}(\mu) = \frac{1-\mu}{\mu K}.\vspace{-2mm} 
\end{equation}

\emph{Proof}: Each transmission can deliver $\mu K$ bits in $1/\log(P)$ channel uses per bit.
\begin{figure}[!t]
\centering
\includegraphics[width=5in]{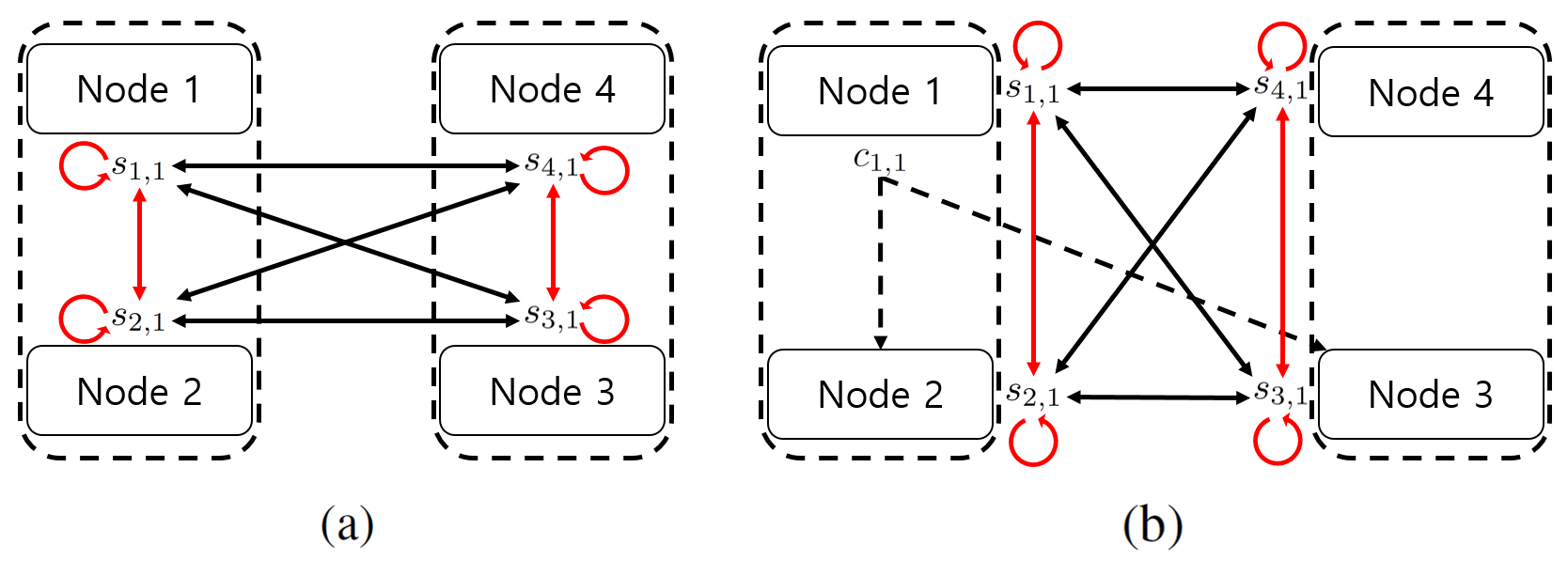}
\caption{Illustration of two transmission schemes for the Shuffle phase with $K = Q = 4$, $\mu = 1/2$: (a) One-shot cooperative ZF precoding [9]; (b) Superposition coding.}
\label{fig_example}
\vspace{-3mm}
\end{figure}
\subsection{One-Shot Cooperative Linear Precoding}
For the wireless model under study, reference \cite{R10} proposed a scheme based on a one-shot cooperative ZF precoding strategy.
This is briefly described next by focusing first on the case $\mu \leq 1/2$.
Assume perfect CSI at the network controller as done in \cite{R10}.
In the Map phase, the files are assigned as explained in Section \ref{subsec:CM}.
In the Shuffle phase, for each block, as shown in Fig. 2(a), the scheme selects $2\mu K$ nodes, which are divided into two clusters of $\mu K$ nodes each.
Since any IV desired by each node is known to $\mu K$ nodes, this selection can be done so that all $\mu K$ nodes in one cluster know the same desired IVs for each node belonging to the other cluster.
Nodes in one cluster transmit $\mu K$ IVs for the $\mu K$ nodes in the other cluster by using cooperative ZF precoding.
The two clusters transmit simultaneously using full-duplex radios.
The $\mu K$ IVs precoded by the nodes in the same cluster can be removed by a node using the available information (red arrows in Fig. 2(a)).
In contrast, the precoded signals from the other cluster are free of inteference thanks to ZF beamforming (black arrow in Fig. 2(a)).
This scheme hence delivers $2\mu K$ sub-IVs through a single transmission at the interference-free rate $\log(P)$.
If $\mu > 1/2$, the maximum number $K$ of sub-IVs can be delivered simultaneously in a similar manner \cite{R10}.

In the presence of outdated CSI at the network controller (but perfect CSI at the receivers), as assumed here, ZF precoding is unable to fully cancel interference and hence the transmission rate needs to be decreased to $\alpha \log(P)$ \cite{R13}.
The NCL of the outline scheme can be obtained as follows. 

\emph{Lemma 2}: For storage capacity $\mu \in \{1/K, \cdots, 1\}$ and $\alpha \in [0,1]$, the NCL of one-shot cooperative linear precoding is given as\vspace{-1mm}
\begin{equation}
\delta_{ZF}(\mu) = \frac{1-\mu}{\alpha K\min(1,2\mu)}.\vspace{-1mm}
\end{equation}

\emph{Proof}: Due to outdated CSI error (3), ZF can deliver $K \min(1,2\mu)$ bits in $1/\alpha\log(P)$ channel uses per bit \cite{R13}.

\section{Superposition coding}
\label{sec:SC}

In this section, we propose a new transmission scheme that applies superposition coding in the Shuffle phase to transmit simultaneously a common message from one node to a subset of other nodes and private messages between clusters of distinct nodes.
The common message is encoded by using coded multicasting from a node to $\mu K$ other nodes, while the private messages are transmitted using cooperative ZF precoding by two clusters of nodes using full-duplex radios.
Decoding takes place sequentially with the common message being decoded first by all receiving nodes.

\emph{Proposition 1}: For storage capacity $\mu \in \{1/K, \cdots, 1\}$ and $\alpha \in [0,1]$, the NCL achieved by superposition based coding is given as
\vspace{-3mm}
\begin{equation}
\delta_{SP}(\mu) = \frac{1-\mu}{(1-\alpha)\mu K + \alpha K \min(1,2\mu)}.
\end{equation}

\emph{Proof}: See Appendix. 

A comparison of the NCL of the proposed superposition based scheme and of the baseline schemes in [4] and [9] described in Section 3 can be found in Fig. 3, where we have set $K = Q = 4$ and $\mu = 1/2$.
It is observed that the superposition based scheme outperforms the baseline schemes in [4] and [9].  
Furthermore, as the CSI precision parameter $\alpha$ grows larger, the NCL of the superposition based scheme increases due to the improved accuracy of ZF precoding.
When $\alpha=0$, superposition reduces to coded multicasting, demonstrating that, with no CSI, it is preferable not to rely on linear precoding.
Conversely when $\alpha = 1$, the NCL of superposition coding coincides with that of ZF precoding scheme of [9].
As a reference, we also show the NCL $\gamma\delta_{CM}(\mu) + (1-\gamma)\delta_{ZF}(\mu)$ obtained via time-sharing, where $0 \leq \gamma \leq 1$ is optimized as a function of $\gamma$.
\begin{figure}[!t]
\centering
\includegraphics[width=4in]{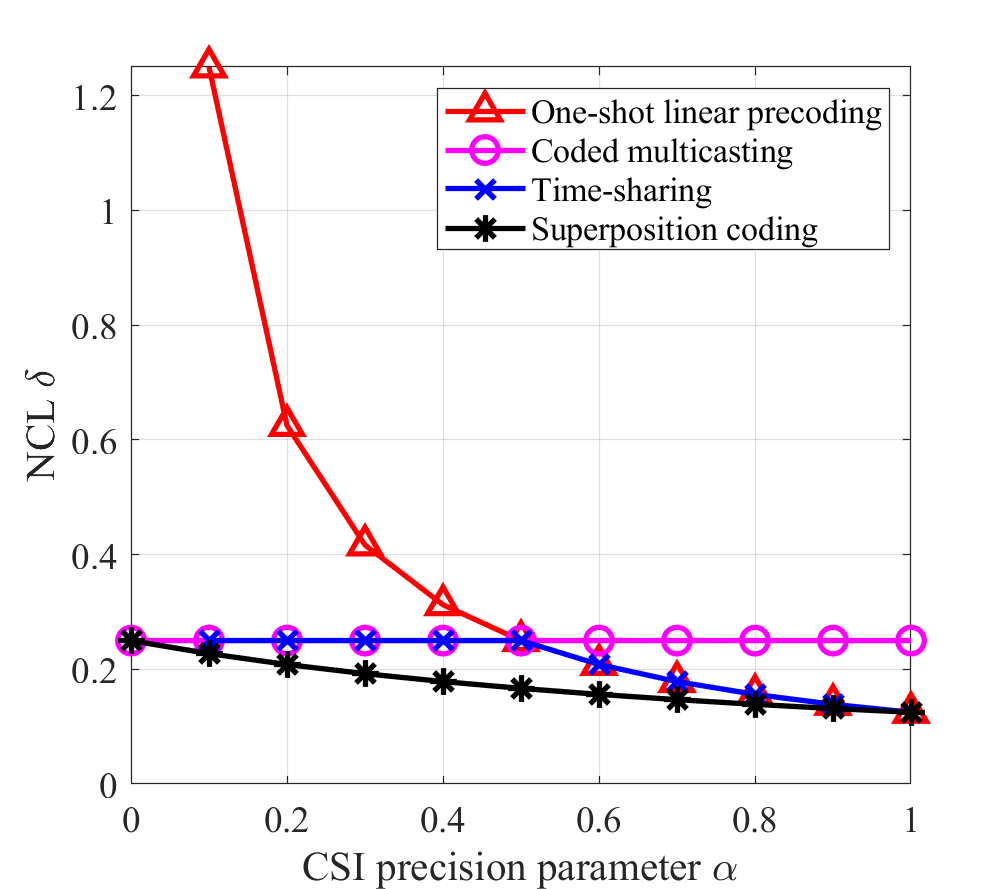}
\caption{NCL $\delta$ as a function of the CSI precision parameter $\alpha$ for the proposed scheme and for the baseline schemes in [4] and [9], with $K = Q = 4$, $\mu = 1/2$.}
\label{fig_sim2}
\end{figure}

\emph{Example}: We present an example to illustrate the operation of the proposed transmission scheme.
Consider $K = 4$ distributed nodes, with $N = 6$ input files, $Q = 4$ output functions, storage capacity $\mu = 1/2$, and CSI accuracy $\alpha = 2/3$. 
In the Map phase, each node $k$ can store $|\mathcal{M}_{k}|=\mu N = 3$ files, which are assigned as $\mathcal{M}_{1} = \{w_{1},w_{2},w_{3}\}$, $\mathcal{M}_{2} = \{w_{1},w_{4},w_{5}\}$, $\mathcal{M}_{3} = \{w_{2},w_{4},w_{6}\}$, and $\mathcal{M}_{4} = \{w_{3},w_{5},w_{6}\}$ as for the two schemes reviewed in Sec. 3.
The output functions assigned to each node are $\mathcal{F}_{1} = \{f_{1}\}$, $\mathcal{F}_{2} = \{f_{2}\}$, $\mathcal{F}_{3} = \{f_{3}\}$, and $\mathcal{F}_{4} = \{f_{4}\}$.

To compute the output function $\mathcal{F}_{k}$ in node $k$, node $k$ has to receive three IVs from other nodes.
The sets of required IVs are $\mathcal{B}_{1} = \{a_{1,4},\hspace{1mm} a_{1,5},\hspace{1mm} a_{1,6}\}$, $\mathcal{B}_{2} = \{a_{2,2},\hspace{1mm} a_{2,3},\hspace{1mm} a_{2,6}\}$, $\mathcal{B}_{3} = \{a_{3,1},\hspace{1mm} a_{3,3},\hspace{1mm} a_{3,5}\}$, and $\mathcal{B}_{4} = \{a_{4,1},\hspace{1mm} a_{4,2},\hspace{1mm} a_{4,4}\}$.
We split each IV into six sub-IVs, two sub-IVs $\{a_{q,n,1}, a_{q,n,2}\}$ with length $F_{CM} = (1-\alpha)F/ [(1-\alpha)\mu K  + 2\alpha \mu K)] = 0.1F$ and four sub-IVs $\{a_{q,n,3}, \cdots, a_{q,n,6}\}$ with length $F_{ZF} = \alpha F/ [(1-\alpha)\mu K  + 2\alpha \mu K)] = 0.2F$.
In the Shuffle phase, each node receives two sub-IVs via coded multicasting and four sub-IVs via ZF precoding.

To this end, transmission in the Shuffle phase takes place over twelve blocks.
In block 1 as shown in Fig. 2(b), all nodes are divided into two clusters of two nodes, with each cluster transmitting two sub-IVs to the other cluster using ZF precoding (solid arrows).
In addition, node 1 transmits an additional coded sub-IV (dashed arrows).
The transmitted signal at node $k$ in block 1 is given as
\begin{displaymath}
x_{k,1} = \left\{ \begin{array}{ll}
c_{k,1} + s_{k,1}, & k = 1,\\
s_{k,1}, &  k \in \{2,3,4\},
\end{array} \right.
\end{displaymath}
where $c_{1,1}$ is the coded sub-IV from node 1 to node 2 and 3 transmitted with power $\mathbb{E}[|c_{1,1}|^{2}] = P-P^{\alpha}$, and $s_{k,1}$ is the ZF-precoded signal transmitted by node $k$ in block 1 with the power  $\mathbb{E}[|s_{k,1}|^{2}] = P^{\alpha}$, for $k \in [4]$. 
The received signal at any node $k$ in block 1 hence is given as
\vspace{-3mm}
\begin{eqnarray}
y_{k,1} \hspace{-3mm}&=&\hspace{-3mm} \sum_{j=1}^{4}h_{j,k}x_{j,1} + n_{k,1}\nonumber\\
&=& \hspace{-3mm}h_{1,k}c_{1,1} + \sum_{j=1}^{4}h_{j,k}s_{j,1} + n_{k,1}.\nonumber
\end{eqnarray}
All nodes decode the coded sub-IV $c_{1,1}$ first, and then decode the desired precoded sub-IV after removing the coded sub-IV by using Successive Interference Cancellation.
In the high-SNR regime, the resulting achievable rates for the coded sub-IV is given as $R_{CM} = \log(P/P^{\alpha}) = (1-\alpha)\log(P)$ due to the interference from the precoded signal.
For the precoded sub-IVs, the achievable rate is instead $R_{ZF} = \alpha\log(P)$.
Therefore, the duration of the block is $F_{CM}/R_{CM} = F_{ZF} / R_{ZF} = 3F/10\log(P)$.
As follows that the NCL of the example is given as $\delta = \frac{12 \times 3F/(10\log(P))}{6 \times 4F/\log(P)} = 0.15$.

\section{Conclusions}
In this paper, we studied a wireless distributed computing system based on the Map-Shuffle-Reduce framework in the presence of imperfect CSI at the network controller.
We proposed a superposition based scheme that simultaneously delivers coded multicasting messages and cooperatively precoded message in the Shuffle phase.  
The superposition based scheme was shown to reduce the normalized communication load as compared to the baseline schemes when CSI is neither perfect nor completely outdated.

\section{Appendix: Proof of Proposition 1}
Here we provide a short description of the generalization of the proposed superposition coding scheme.
In the Map phase, the input files are assigned using the same method as in Section \ref{subsec:CM}.
Extending the example in Section \ref{sec:SC}, we split each IV into $\mu K + \min (K, 2\mu K)$ sub-IVs.
The first $\mu K$ sub-IVs are of length $F_{c} = F (1-\alpha) / [(1-\alpha)\mu K  + \alpha K\min (1, 2\mu )]$ bits, while the length of remaining $\min(K,2\mu K)$ sub-IVs is $F_{s} = F\alpha / [(1-\alpha)\mu K  + \alpha K\min (1, 2\mu )]$.
For each block, $\min (K, 2\mu K)$ nodes transmit precoded signals and one of the nodes transmits a coded sub-IV.
The precoded signals are transmitted by dividing the set of active nodes into two equal clusters as in the scheme of \cite{R10}.
As in the example, the transmit power of coded sub-IV and precoded sub-IV are set to $P$ and $P^{\alpha}$, respectively.
Each of the $\min(K,2\mu K)$ nodes decodes the coded sub-IV first in the decoding process and then decodes the precoded sub-IV.
Since the received SNR of coded sub-IV and precoded sub-IV are $P^{1-\alpha}$ and $P^{\alpha}$ respectively, the respective achievable rates are $(1-\alpha)\log(P)$ and $\alpha\log(P)$, and the duration of each block is $F/[ (1-\alpha)\mu K + \alpha K\min(1,2\mu)]\log(P)$.
With this scheme, $\mu K$ coded sub-IVs and $\min(K,2\mu K)$ precoded sub-IVs can be delivered through a single transmission.
Since the number of coded sub-IVs and precoded sub-IVs to be transmitted and $NQ(1-\mu)\mu K$ and $NQ(1-\mu)\min(K,2 \mu K)$, respectively, a total of $NQ(1-\mu)$ blocks are required to transmit all IVs. 
As a result, the NCL of the proposed superposition coding scheme is given as
\begin{equation}
\delta_{SP}(\mu)= \frac{NQ(1-\mu) \times F/[ (1-\alpha)\mu K + \alpha K\min(1,2\mu)]\log(P)}{NQF/\log(P)} = \frac{1-\mu}{(1-\alpha)\mu K + \alpha K \min(1,2\mu)}.
\end{equation}

\section*{Acknowledgment}
The work of S. Ha and J. Kang was supported by the National Research Foundation of Korea (NRF) grant funded by the Korea government (MSIT) (No. 2017R1A2B2012698). J. Zhang and O. Simeone have received funding from the European Research Council (ERC) under the European Union’s Horizon 2020 Research and Innovation Programme (Grant Agreement No. 725731). The work of O. Simeone was also funded in part by the U.S. National Science Foundation under Grant 1525629.

\bibliographystyle{IEEEbib}
\bibliography{refs}

\end{document}